# Element Replacement Approach by Reaction with Lewis Acidic Molten Salts to Synthesize Nanolaminated MAX Phases and MXenes

Mian Li[1], Jun Lu[2], Kan Luo[1], Youbing Li[1], Keke Chang[1], Ke Chen[1], Jie Zhou[2], Johanna Rosen[2], Lars Hultman[2], Per Eklund[2], Per O.Å. Persson[2], Shiyu Du[1], Zhifang Chai[1], Zhengren Huang[1], Qing Huang[1*]

(1. Engineering Laboratory of Advanced Energy Materials, Ningbo Institute of Industrial Technology, Chinese Academy of Sciences, Ningbo, Zhejiang, 315201, China; 2. Department of Physics, Chemistry, and Biology (IFM), Linköping University, 58183 Linköping, Sweden)

Corresponding to: huangqing@nimte.ac.cn

**Abstract**

Nanolaminated materials are important because of their exceptional properties and wide range of applications. Here, we demonstrate a general approach to synthesize a series of Zn-based MAX phases and Cl-terminated MXenes originating from the replacement reaction between the MAX phase and the late transition metal halides. The approach is a top-down route that enables the late transitional element atom (Zn in the present case) to occupy the A site in the pre-existing MAX phase structure. Using this replacement reaction between Zn element from molten $ZnCl_2$ and Al element in MAX phase precursors ($Ti_3AlC_2$, $Ti_2AlC$, $Ti_2AlN$, and $V_2AlC$), novel MAX phases $Ti_3ZnC_2$, $Ti_2ZnC$, $Ti_2ZnN$, and $V_2ZnC$ were synthesized. When employing excess $ZnCl_2$, Cl terminated MXenes (such as $Ti_3C_2Cl_2$ and $Ti_2CCl_2$) were derived by a subsequent exfoliation of $Ti_3ZnC_2$ and $Ti_2ZnC$ due to the strong Lewis acidity of molten $ZnCl_2$. These results indicate that A-site element replacement in traditional MAX phases by late transition metal halides opens the door to explore MAX phases that are not thermodynamically stable at high temperature and would be difficult to synthesize through the commonly employed powder metallurgy approach. In addition, this is the first time that exclusively Cl-terminated MXenes were obtained, and the etching effect of Lewis acid in molten salts provides a green and viable route to prepare MXenes through an HF-free chemical approach.

**Introduction**

The family of nanolaminates called MAX phases, and their two-dimensional (2D) derivative MXenes, is attracting significant attention owing to their unparalleled properties.[1-7] The MAX phases have the formula of $M_{n+1}AX_n$ (n =1−3), where M is an early transition metal, A is an element traditionally from groups 13-16, X is carbon or nitrogen. The unit cell of MAX phases is comprised of $M_6X$ octahedral (e.g. $Ti_6C$) interleaved with layers of A elements (e.g., Al). When etching the A-site atoms by HF or other acids, the retained $M_{n+1}X_n$ sheets form 2D sheets, called MXenes. Theses 2D derivatives show great promise for applications as battery electrodes, supercapacitors, electromagnetic absorbing and shielding coating, catalyst and carbon capture.[8-15]

So far, about 80 ternary MAX phases have been experimentally synthesized, with more continuously being studied, often guided by theory.[16, 17] However, MAX phases with the A-site elements of late transition metal (e.g., Fe, Ni, Zn, and Pt), which are expected to exhibit diverse functional properties (e.g., magnetism and catalysis), are difficult to synthesize. For these late transition metals, their M-A intermetallics are usually more stable than corresponding MAX phases at high synthesis temperatures, that means the target MAX phases can hardly be achieved by a thermodynamic equilibrium process such as hot pressing (HP) and spark plasma sintering (SPS).

In 2017, Fashandi *et al.* synthesized MAX phases with noble-metal elements in the A site, obtained through a replacement reaction.[18, 19] The replacement reaction was achieved by replacement of Si by Au in the A layer of $Ti_3SiC_2$ at high annealing temperature with a thermodynamic driving force for separation of Au and Si at moderate temperature, as determined from the Au-Si binary phase diagram. The formation of $Ti_3AuC_2$ is driven by an A-layer diffusion process. Its formation is preferred over competing phases (e.g. Au-Ti alloys), and the MAX phase can be obtained at a moderate temperature. Similarly, Yang *et al.* also synthesized $Ti_3SnC_2$ by a replacement reaction between Al atom in $Ti_3AlC_2$ and $SnO_2$,[20] although $Ti_3SnC_2$ can also be synthesized by a hot isostatic pressing (HIP) route.[21] Their work implies the feasibility of synthesis of novel MAX phases through replacement reactions.

It is worth noting that the synthesis of MAX phases by the A-site element exchange approach is similar to the preparation of MXenes by A-site element etching process. Both are top-down routes that make modification on the A atom layer of pre-existing structure of MAX phase, which involve the extraction of A-site atoms and intercalation of new species (e.g. metallic atoms or functional terminals) at particular lattice position. Based on this idea, we introduce here a general approach to synthesize a series of novel nanolaminated MAX phases and MXenes based on the element exchange approach in the A-layer of traditional MAX phase. The late transition metal halides, e.g. $ZnCl_2$ in this study, are so-called Lewis acids in their molten state.[22-25] These molten salts can produce strong electron-accepting ligands, which can thermodynamically react with the A element in the MAX phases. Simultaneously, certain types of atoms or ions can diffuse into the two-dimensional atomic plane and bond with the unsaturated $M_{n+1}X_n$ sheet to form corresponding MAX phases or MXenes. The flexible selection of salt constituents can provide sufficient room to control the reaction temperature and the type of intercalated ions. In the present work, a variety of novel MAX phase ($Ti_3ZnC_2$, $Ti_2ZnC$, $Ti_2ZnN$, and $V_2ZnC$) and MXenes ($Ti_3C_2Cl_2$ and $Ti_2CCl_2$) were synthesized by elemental replacement in the A atomic plane of traditional MAX phases in $ZnCl_2$ molten salts. The results indicate a general and controllable approach to synthesize novel nanolaminated MAX phases and the derivation of halide group-terminated MXenes from its respective parent MAX phase.

**Experimental and Computational Details**

**Preparation of MAX phases and MXenes.** All the obtained new MAX phases ($Ti_3ZnC_2$, $Ti_2ZnC$, $Ti_2ZnN$ and $V_2ZnC$, denoted in brief as Zn-MAX) and MXenes ($Ti_3C_2Cl_2$ and $Ti_2CCl_2$, denoted in brief as Cl-MXenes) were prepared by a reaction between the traditional MAX phase precursors ($Ti_3AlC_2$, $Ti_2AlC$, $Ti_2AlN$, and $V_2AlC$, denoted as Al-MAX) and $ZnCl_2$. For synthesizing the Zn-MAX, a mixture powder composed of the Al-MAX: $ZnCl_2$=1:1.5 (molar ratio) was used as the starting material. For synthesizing the Cl-MXenes, a mixture powder composed of Al-MAX:$ZnCl_2$=1:6 (molar ratio) was used as the starting material.

The starting material was mixed thoroughly using a mortar under the protection of nitrogen in a glovebox. Then the as-obtained mixture powder was taken out from the glove box and placed into an alumina crucible. The alumina crucible was loaded into a tube furnace and heat-treated at 550 °C for 5 h under the protection of Ar gas. After the reaction, the product was washed by de-ionized water to remove the residual $ZnCl_2$ and the final product was dried at 40 °C. Finally, the target reaction product was obtained. The weight change during various steps of the process was provided in the supporting information (Table S1, Table S2, and Fig. S1).

In addition, the starting materials of $Ti_3AlC_2$:$ZnCl_2$=1:1 to 1:6 were heat-treated at 550 °C for 5 h to investigate the influence of $Ti_3AlC_2$/$ZnCl_2$ ratio on the composition of product. The starting materials of $Ti_3AlC_2$:$ZnCl_2$=1:6 was heat-treated at 550 °C for different time (0.5 h to 5h) to investigate the phase evolution of the reaction product.

**Preparation of MAX phase precursors.** The $Ti_3AlC_2$, $Ti_2AlC$, $Ti_2AlN$, and $V_2AlC$ MAX phase precursors were powders prepared by a molten-salt method as was reported previously.[14, 26, 27] Metal carbides/nitrides powders (TiC, TiN, and VC), M-site metal powders (Ti and V) and A-site metal powders (Al) were used as the source for synthesizing the target MAX phases. NaCl and KCl with the molar ratio of 1:1 was used as the source for the molten salt bath. All these metal carbides/nitrides were ~10 μm in particle size and were purchased from Pantian Nano Materials Co. Ltd. (Shanghai, China). Ti, V, and Al powders were 300 mesh in particle size and were purchased from Yunfu Nanotech Co. Ltd. (Shanghai, China). NaCl and KCl at analytical grade were purchased from Aladdin Industrial Co. Ltd. (Shanghai, China).

The starting materials, with the composition shown in Table 1, were mixed in a mortar and placed in an alumina crucible. Under the protection of argon, the alumina crucible was packed in a tube furnace and heat-treated to 1100 °C with the rate of 4 °C/min and kept 3 h to accomplish the reaction. After reaction, the tube furnace was cooled to room temperature with the rate of 4 °C/min. Then the product was washed by deionized water to remove the NaCl and KCl, and the residual product was dried at 60 °C. As a result, the target MAX phase precursors were obtained.

Table 1. The composition of starting materials for synthesizing the Al-MAX phases.

| MAX phase | Composition of starting materials |
| --- | --- |
| $Ti_3AlC_2$ | TiC : Ti : Al : NaCl : KCl = 2 : 1 : 1 .1 : 4: 4 |
| $Ti_2AlC$ | TiC : Ti : Al : NaCl : KCl = 1 : 1 : 1 .1 : 4: 4 |
| $Ti_2AlN$ | TiN : Ti : Al : NaCl : KCl = 1 : 1 : 1 .1 : 4: 4 |
| $V_2AlC$ | VC : V : Al : NaCl : KCl = 1 : 1 : 1 .1 : 4: 4 |

**Characterization.** Scanning electron microscopy (SEM) was performed in a thermal field emission scanning electron microscope (Thermo scientific, Verios G4 UC) equipped with an energy-dispersive spectroscopy (EDS) system. Transmission electron microscopy (TEM) and scanning transmission electron microscopy (STEM) was performed in the Linköping monochromated, high-brightness, double-corrected FEI Titan³ 60–300 operated at 300 kV, equipped with a SuperX EDS system. X-ray diffraction (XRD) analysis of the composite powders was performed using a Bruker

D8 ADVANCE X-ray diffractometer with Cu Kα radiation at a scan rate of 2 °/min. X-ray photoelectron spectra (XPS) were performed in an XPS system (Axis Ultra DLD, Kratos, UK) with a monochromatic Al X-ray source. The binding energy (BE) scale were assigned by adjusting the C 1s peak at 284.8 eV.

**Computational Methods.** First-principles density-functional theory (DFT) calculations were carried out using the CASTEP module.[28, 29] A plane wave cutoff of 500 eV and the ultrasoft pseudopotentials in reciprocal space with the exchange and correlation effects represented by generalized gradient approximation (GGA) of the Perdew-Breke-Ernzerhof (PBE) functional were employed in the structure calculations.[30-33] The atomic positions were optimized to converge towards the total energy change smaller than $5\times10^{-6}$ eV/atom, maximum force over each atom below 0.001 eV/Å, pressure smaller than 0.001 GPa, and maximum atomic displacement not exceeding $5\times10^{-4}$ Å. Phonon calculations were also performed by the finite displacement approach implemented in CASTEP to evaluate the dynamically stability.[34, 35] The cleavage energies $E$ of different M and Zn atomic layer in $Ti_3ZnC_2$, $Ti_2ZnC$, $Ti_2ZnN$, and $V_2ZnC$ were calculated using the equation $E=(E_{broken}-E_{bulk})/2A$,[36] where $E_{bulk}$ and $E_{broken}$ are the total energies of bulk MAX and the cleaving structures with a 10 Å vacuum separation in the corresponding M and Zn atomic layers, $A$ is the cross-sectional surface area of the MAX phase materials.

The CALPHAD approach was applied to calculate the phase diagrams of the Ti-Al-C and Ti-Zn-C systems. The Ti-Al-C system has been well assessed by Witusiewicz et al.[37] and the thermodynamic dataset were adopted in this work. Due to lack of experimental data on the ternary Ti-Zn-C compounds, first-principles calculations were conducted to support the CALPHAD work.[38] The formation enthalpies of the $Ti_2ZnC$ and $Ti_3ZnC_2$ ternary compounds were computed to be -414.63 and -230.16 kJ/mol, respectively. Their Gibbs free energy functions were then determined with the Neumann-Kopp rule and added in the CALPHAD-type dataset of the Ti-Zn-C system, which included the thermodynamic parameters of the binary Ti-Zn and Zn-C systems.[39] With the established thermodynamic dataset, the isothermal sections of the Ti-Al-C and Ti-Zn-C systems at 1300 and 550 °C were computed. All the calculations were performed with the Thermo-Calc software.

**Results and discussion**

**Zn-MAX phases.** $Ti_3ZnC_2$ was prepared by using the starting materials of $Ti_3AlC_2$ and $ZnCl_2$ with mole ratio of 1:1.5. Fig.1a shows the XRD patterns of the initial phase $Ti_3AlC_2$ and the final product $Ti_3ZnC_2$. Compared to $Ti_3AlC_2$, the XRD peaks of $Ti_3ZnC_2$, e.g., (103), (104), (105), and (000$l$) peaks, are shifted towards lower angles, indicating a larger lattice constant caused by the replacement of Al atoms by Zn atoms. Note that the relative intensity of (0004), (0006) peaks increased while that of (0002) peaks decreased. This is caused by the change in structure factor by the replacement of the A atoms. According to a Rietveld refinement of the XRD pattern (Fig. S2), the determined $a$ and $c$ lattice parameter of $Ti_3ZnC_2$ are 3.0937 Å and 18.7206 Å, which are larger than that of $Ti_3AlC_2$ ($a$=3.080 Å, $c$=18.415 Å).[3]

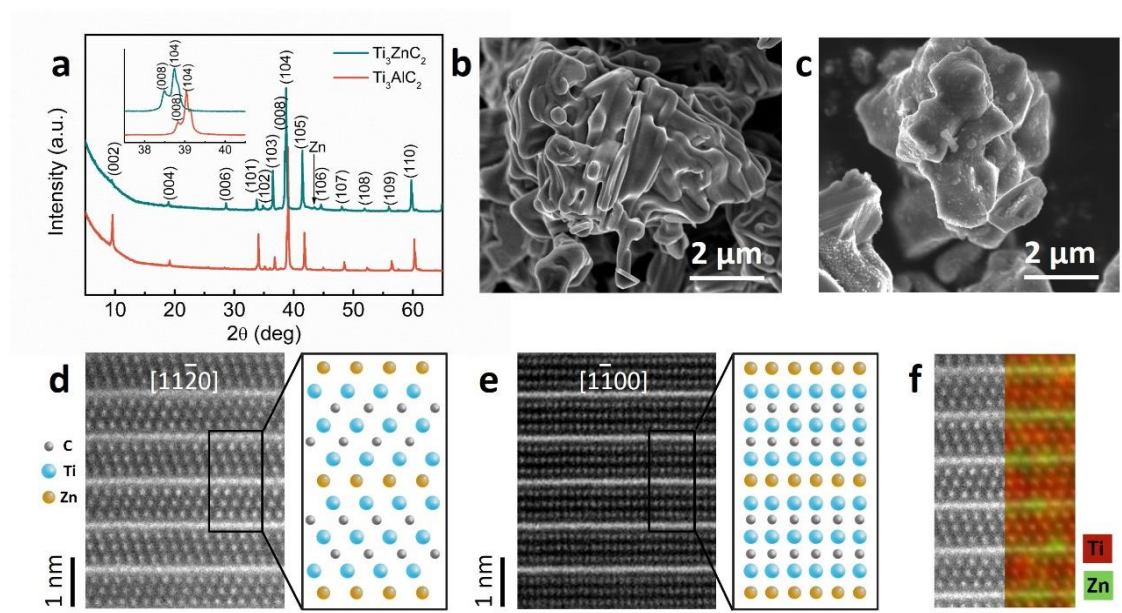

Fig.1 Characterization of $Ti_3ZnC_2$. (a) XRD patterns of $Ti_3AlC_2$ and $Ti_3ZnC_2$. (b) SEM image of $Ti_3AlC_2$. (c) SEM image of $Ti_3ZnC_2$. (d), (e) High-resolution (HR)-STEM image of $Ti_3ZnC_2$, showing the atomic positions from different orientations. (f) HR-STEM and the corresponding EDS map of $Ti_3ZnC_2$.

Fig. 1b and c shows SEM images of $Ti_3AlC_2$ and $Ti_3ZnC_2$ powders. The $Ti_3AlC_2$ precursor exhibits the typical layered structure of MAX phases. In contrast, the layered structure of $Ti_3ZnC_2$ becomes less distinct, which is possibly attributed to the dissolution of the edges of the powders in the molten salt. EDS analysis (Fig. S3) indicates that the elemental composition of $Ti_3ZnC_2$ is Ti: Zn: C= 50.2:16.1:26.9 (atomic ratio), and a small amount of Al (2.4 at.%) and O (3.4 at.%). This observation indicates that a nearly complete replacement of Al by Zn was achieved.

Atomically-resolved STEM in combination with lattice resolved EDS was employed to demonstrate the elemental organization of $Ti_3ZnC_2$. Fig. 1d and 1e shows the atomic projections with the electron beam along $[11\bar{2}0]$ and $[1\bar{1}00]$, respectively. $Ti_3C_2$ sublayers exhibit the characteristic zig-zag pattern of MAX phases when observed along $[11\bar{2}0]$. Atomically thin layers of Zn, which appear brighter due to mass-contrast imaging, separate the $Ti_3C_2$ sheets. Fig. 1f shows a STEM image with the lattice resolved elemental map. Note that the individual Zn atoms in these images cannot readily be distinguished, as they form a continuous line. This observation may be explained by in-plane vibrations of the Zn atoms, presumably caused by weaker bonding to the $Ti_3C_2$ layers than in the $Ti_3AlC_2$ case.

Fig. 2 shows the phase identification of products derived from $M_2AX$ precursors ($Ti_2AlC$, $T_2AlN$ and $V_2AlC$ were studied here). The XRD patterns in Fig.2a exhibit a similar variation as that of the $M_3AX_2$ system, (103), (106), and (110) peaks shifted towards lower angles and relative intensity of (000$l$) peaks changed. The STEM images (Fig. 2b-c) and the EDS analysis (Fig. S4) further confirmed the formation of the corresponding $Zn-M_2AX$ ($T_2ZnC$, $Ti_2ZnN$ and $V_2ZnC$) after the reactions.

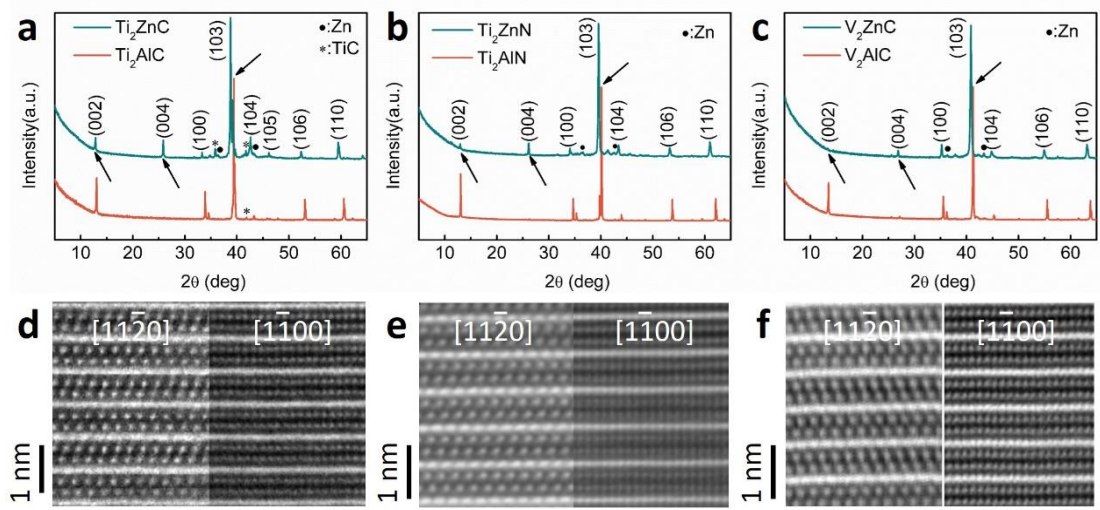

Fig.2 Characterization of $Ti_2ZnC$, $Ti_2ZnN$, and $V_2ZnC$. (a)-(c) XRD patterns showing the three MAX phases with their respective precursors, the arrows indicate the difference. (d)-(f) HR-STEM image of $Ti_2ZnC$, $Ti_2ZnN$, and $V_2ZnC$, showing the atomic positions in different orientations.

**Cl-MXenes.** A mixture of $Ti_3C_2Cl_2$ MXene sheets and Zn spheres were obtained by using the starting materials of $Ti_3AlC_2$ and $ZnCl_2$ with mole ratio of 1:6. The SEM image (Fig. 3a) and TEM images (Fig. S5) of $Ti_3C_2Cl_2$ shows exfoliation along the basal planes. The corresponding EDS analysis (Fig. S6) indicates that the elemental composition of the MXene is Ti:C:Cl =43.2:21.5:25.3 in atomic ratio, and small amount of Zn (0.7 at.%), Al (2.9 at.%) and O(6.3 at.%). The presence of oxygen is reasonable due to the prevailing O-containing compounds like $Al(OH)_3$, which is the hydrolysis product of $AlCl_3$. Note that our theoretical calculation results indicate that the Cl terminations can strongly bond on the MXene surfaces, while not competitive to O-containing terminals.[40] Thus, a small part of Cl terminations might be replaced by O-containing terminals during the producing process like water washing, which could also contribute to the oxygen element detected on the surface. In addition, large Zn spheres can also be observed in the product, which can be easily distinguished from the MXenes (Fig. S7).

Fig. 3b shows the XRD patterns of the as-reacted and HCl-rinsed products. Besides diffraction peaks characteristic of Zn metal, the XRD patterns of the as-reacted product is similar to that of the $Ti_3C_2$ MXene produced by the commonly used HF etching method. Most of the non-basal-plane peaks of $Ti_3AlC_2$, e.g. the intense (104) peak at ~39°, weakened significantly or disappeared. The (0002) peaks downshifts to a lower angle of $2\theta=7.94$°C, attributed to an increased $c$ lattice parameter of 22.24 Å. This $c$ lattice parameter agrees with the theoretical value calculated by DFT (22.34 Å), and is larger than that of the $Ti_3C_2T_x$ produced by HF etching (19.49 Å for $Ti_3C_2(OH)_2$ and 21.541 Å for $Ti_3C_2F_2$).[3] Note that unlike the typical broad (000$l$) peaks of $Ti_3C_2T_x$ produced by HF etching, the (000$l$) peaks of $Ti_3C_2Cl_2$ are sharp and intense, indicating an ordered crystal structure. The etching mechanism of MAX phase in molten salts may be of great interest since the chemical process is considerably safer and cleaner compared to the HF etching method. Regarding the etching chemistry, the reduction of Zn cation in the molten salts is much similar to the generation of $H_2$ in the HF solution to etch $Ti_3AlC_2$.[3]

The as-reacted product was treated by a 5 wt.% HCl solution at 25 °C for 2h and washed by de-ionized water to remove Zn according to reaction 1.

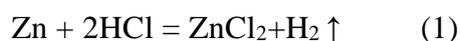
$$Zn + 2HCl = ZnCl_2 + H_2 \uparrow \qquad (1)$$

After HCl treating, the XRD peaks corresponding to Zn disappeared, while the peaks corresponding to $Ti_3C_2Cl_2$ remains unchanged. The SEM and EDS analysis also confirmed that the morphology and composition of $Ti_3C_2Cl_2$ was not influenced by the HCl treatment (Fig. S8), indicating that $Ti_3C_2Cl_2$ is stable in the HCl solution.

A STEM/EDS study was further employed to investigate the as-synthesized MXenes. As shown in Fig. 3c, Cl atoms terminates the surface of $Ti_3C_2$. Note that the $Ti_3C_2Cl_2$ MXenes show good ordering along the basal planes, which is in agreement with the intense (000$l$) peaks in the XRD patterns. For a more detailed investigation on the Cl terminations, please see **ref 40**.[40]

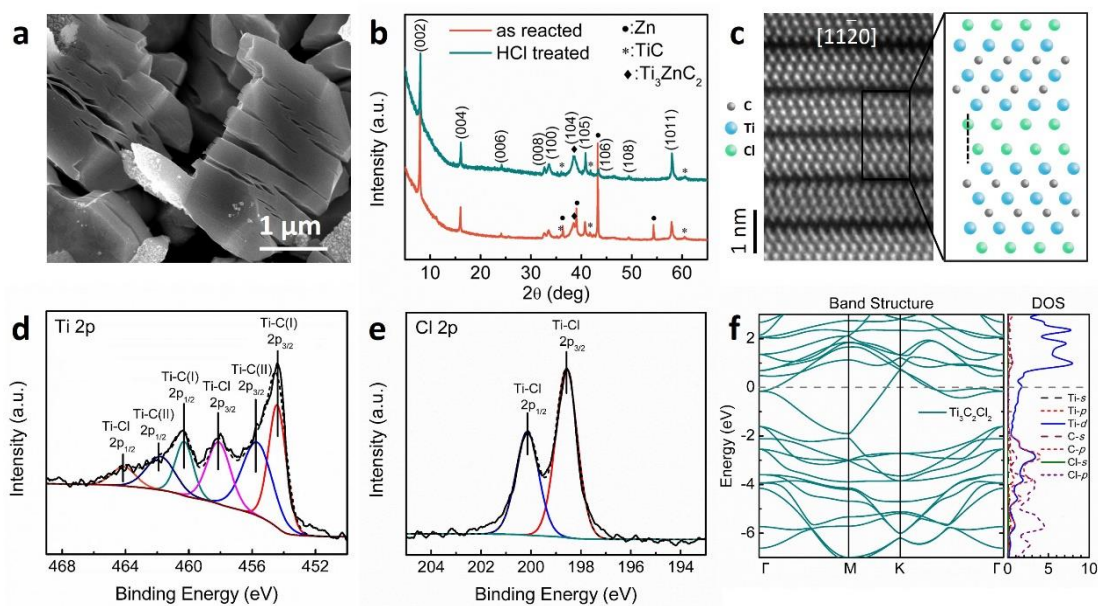

Fig.3 Characterization of $Ti_3C_2Cl_2$. (a) XRD patterns from the as-reacted product and HCl-treated product. (b) SEM image showing $Ti_3C_2Cl_2$ in the as-reacted sample. (c) HR-STEM image showing the atomic positions of $Ti_3C_2Cl_2$. (d) Ti 2p XPS analysis of the as-reacted product. (e) Cl 2p XPS analysis of the as-reacted product. (f) Band structure of $Ti_3C_2Cl_2$.

A Ti 2p X-ray photoelectron spectrum of the as-reacted sample was shown in Fig.3d. The peak at 454.4 eV and 455.7 eV are assigned to the Ti-C (I) ($2p_{3/2}$) and Ti-C (II) ($2p_{3/2}$) bond.[41, 42] The peak at 458.1 eV, attributing to a high valence Ti compound, is assigned to the Ti-Cl ($2p_{3/2}$) bond.[43, 44] Besides, the peaks at 460.3 eV, 461.8 eV and 464.1 eV are assigned to the Ti-C (I) ($2p_{1/2}$), Ti-C (II) ($2p_{1/2}$), and Ti-Cl ($2p_{1/2}$) bonds, respectively. Fig. 3e is the Cl 2p spectrum. The peaks at 198.6 eV and 200.1 eV agrees well with the position of Cl-Ti ($2p_{1/2}$) and Cl-Ti ($2p_{3/2}$) bonds[43, 44] which confirmed the presence of Ti-Cl bonds. The Ti:Cl ratio determined by the XPS analysis is 2.94:2, which agrees well with the EDS and STEM results. Detailed information on the XPS analysis was provided in upporting information (Fig. S9, Fig S10 and Table S4).

The electronic band structures and phonon spectra of single-layer $Ti_3C_2Cl_2$ calculated by DFT is shown in Fig. 2f. The layer is metallic in nature with a finite density of states at the Fermi level. The phonon spectra show that all phonon frequencies are positive (Fig S11), i.e., $Ti_3C_2Cl_2$ MXene is dynamically stable.

Similar to the production of $Ti_3C_2Cl_2$, $Ti_2CCl_2$ was also obtained by using $Ti_2AlC$ and $ZnCl_2$ with mole ratio of 1:6 as the staring materials. Characterization is provided in supporting information (Fig. S12). The phonon spectra of $Ti_2CCl_2$ show that all phonon frequencies are positive (supporting information, Fig. S13a). The calculated band structure indicates that $Ti_2CCl_2$ has a metallic nature, similar to $Ti_3C_2Cl_2$ (Fig. S13b).

In contrast, for the starting materials $Ti_2AlN:ZnCl_2=1:6$ and $V_2AlC:ZnCl_2=1:6$, their final reaction product remained $Ti_2ZnN$ and $V_2ZnC$, but not the corresponding Cl-MXenes (Fig. S14).

**Formation Mechanism.** Taking $Ti_3ZnC_2$ as example, the formation of the Zn-MAX phase can be stated as the following simplified reaction:

$$Ti_3AlC_2 + 1.5ZnCl_2 = Ti_3ZnC_2 + 0.5Zn + AlCl_3 \uparrow \quad (2)$$

$$Ti_3AlC_2 + 1.5ZnCl_2 = Ti_3C_2 + 1.5Zn + AlCl_3 \uparrow \quad (3)$$

$$Ti_3C_2 + Zn = Ti_3ZnC_2 \quad (4)$$

Reaction (2) is the general reaction that can be divided into two sub-reactions as reaction (3)-(4). It is well known that $ZnCl_2$ has a melting pointing of ~280 °C and is ionized into $Zn^{2+}$ and $ZnCl_4^{2-}$ tetrahedron in its molten state.[23, 24] The coordinately unsaturated $Zn^{2+}$ is a strong acceptor for $Cl^-$ and electrons, which act as the Lewis acidity in $ZnCl_2$ molten salt. In this acidic environment, the weakly bonded Al atoms in $Ti_3AlC_2$ can be easily converted into $Al^{3+}$ by a redox reaction (reaction 3). The as-produced $Al^{3+}$ would further bond with the $Cl^-$ to form $AlCl_3$, which has a boiling point of ~180 °C and is expected to rapidly evaporate at the reaction temperature (550 °C). Meanwhile, the *in-situ* reduced Zn atoms intercalate into the $Ti_3C_2$ layers and fill the A-sites of the MAX phase previously occupied by Al atoms, resulting in the formation of $Ti_3ZnC_2$. The evaporation of $AlCl_3$ provides the driving force for the out-diffusion of the Al atom, while the A-site vacancies in the MAX phase enable the Zn atoms to intercalate into the $Ti_3C_2$ layers. In fact, the Al-Zn system forms a eutectic alloy at ~385 °C, well below the reaction temperature. The formation of liquid Al-Zn eutectic in the two-dimensional A-sublayer of MAX phase, together with liquid state ZnCl$_2$, also accelerate the in-plane diffusion of Al and Zn atoms. Actually, the calculated Ti-Zn-C phase diagram indicates that Ti$_3$ZnC$_2$ is not stable at 1300 °C due to the existence of competitive phases (Fig. S15a). This is the main reason why the Zn-MAX phases are unavailable through traditional solid reaction methods. Otherwise, the main composition at high temperature should be Ti-Zn alloy and TiC phase. However, the calculated phase diagram at 550 °C suggests that Ti$_3$ZnC$_2$ is thermodynamically competitive to the Ti-Zn alloys and should be equilibrium at this temperature (Fig. S15b). In contrast, the well-known stable Al-MAX phase exists in the Ti-Al-C phase diagrams both at 1300 °C and 550 °C (Fig. S15c and d). The A-site replacement approach take advantage of diffusion of elements of interest at low temperatures and avoid the reaction to form competitive M-A alloys at high temperature that are usually required to generate the MX sublayer of MAX phase.

Although the formation mechanism of Zn-MAX phases is reasonably explained as above, the formation process of Cl-MXenes remains an open question. It is clear that the molar ratio of Al-MAX:ZnCl$_2$ plays a key role in determining the final product. With the increasing ZnCl$_2$ ratio in the starting materials, the final product gradually transformed from Ti$_3$ZnC$_2$ to Ti$_3$C$_2$Cl$_2$. The XRD patterns of the reaction products for the starting materials with different Ti$_3$AlC$_2$:ZnCl$_2$ ratios (Fig. S16) clearly illustrates this phase evolution. Based on these conditions we propose that the formation of Ti$_3$C$_2$Cl$_2$ follows a two-step reaction: 1) the formation of Ti$_3$ZnC$_2$ in an initial stage, and 2) the further etching of Ti$_3$ZnC$_2$ in excess ZnCl$_2$ melt to form Ti$_3$C$_2$Cl$_2$.

In order to confirm this hypothesis, the starting materials of Ti$_3$AlC$_2$ and ZnCl$_2$ with molar ratio of 1:6 were heat-treated at 550 °C for different annealing times to show the phase evolution of the reaction product. As shown in Fig. 4a, the reaction product shows a variation tendency from Ti$_3$AlC$_2$ to Ti$_3$ZnC$_2$ and next to Ti$_3$C$_2$Cl$_2$ with increasing reaction time. The product after 0.5 h annealing remained Ti$_3$AlC$_2$, while after 1.0 h, it has converted into Ti$_3$ZnC$_2$. Note that for the 1.0 h sample, the (000$l$) peaks of Ti$_3$ZnC$_2$ are less obvious than that in Fig.1a, indicating that the crystal ordering along (000$l$) plane was reduced. This is because the fact that part of Zn atoms were already extracted from the A-layer. When the reaction time was further increased to 1.5 h, the non-basal-plane peaks of Ti$_3$ZnC$_2$ nearly diminished, and the (000$l$) of Ti$_3$C$_2$Cl$_2$ emerged. Note that intense peaks of Zn were also observed in this sample, indicating that the formation of $Ti_3C_2Cl_2$ is accompanied by the generation of free Zn. The 3.0 h sample is the same as the result in Fig.3, a mixture of highly crystallized $Ti_3C_2Cl_2$ and Zn were obtained. The phase evolution was further confirmed by the EDS mapping analysis of the 1.5 h sample. As shown in Fig. 4b, this sample shows a core-shell structure, in which the core region is dense and rich of Zn, and the edge region is delaminated along the in-plane direction and rich of Cl. This core-shell microstructure well shows an intermediate state of the conversion from $Ti_3ZnC_2$ to $Ti_3C_2Cl_2$.

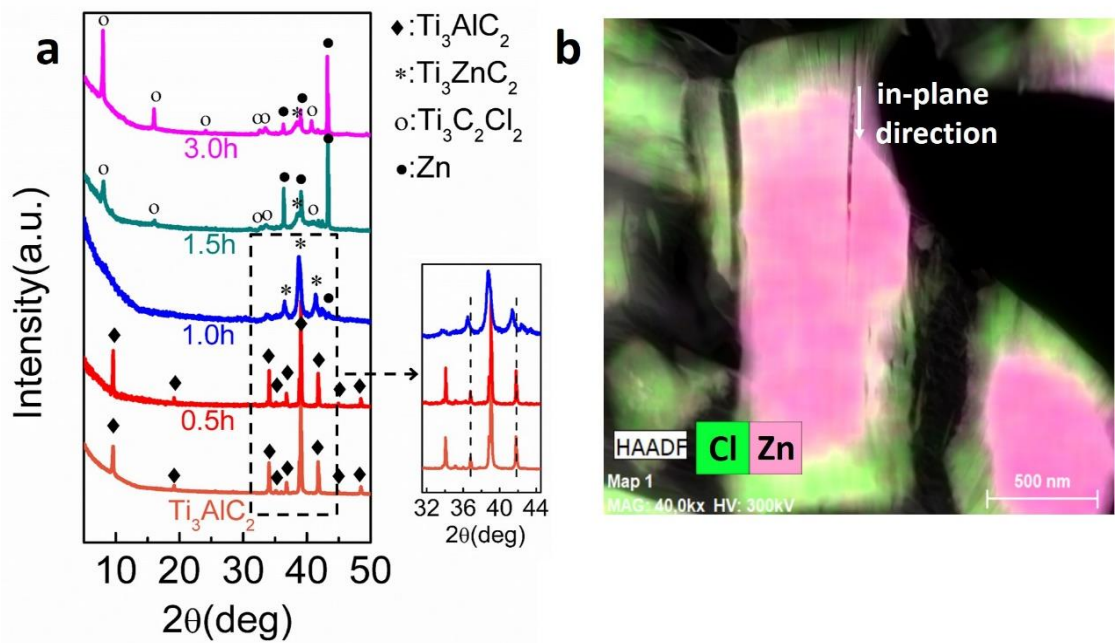

Fig.4 Phase evolution. (a) XRD patterns of the product of $Ti_3AlC_2$:$ZnCl_2$=1:6 system with different reaction time, showing the phases evolution from $Ti_3AlC_2$ to$Ti_3ZnC_2$ to $Ti_3C_2Cl_2$.(b) EDS mapping analysis of the 1.5h sample, showing the $Ti_3ZnC_2$@$Ti_3C_2Cl_2$ core-shell structure.

The above results confirmed that $Ti_3C_2Cl_2$ was the exfoliation product of $Ti_3ZnC_2$, not directly derived from $Ti_3AlC_2$. This process can be reasonably described by the following simplified equations:

$$Ti_3ZnC_2 + ZnCl_2 = Ti_3C_2Cl_2 + 2Zn \quad (5)$$
$$Ti_3ZnC_2 + Zn^{2+} = Ti_3C_2 + Zn_2^{2+} \quad (6)$$
$$Ti_3C_2 + 2Cl^- = Ti_3C_2Cl_2 + 2e^- \quad (7)$$
$$Zn_2^{2+} + 2e^- = 2Zn \quad (8)$$

Reaction (5) is the general reaction that can be divided into three sub-reactions as reactions (6)-(8). Reaction (6) can be explained by the dissolution of Zn in molten

ZnCl$_2$. Previous research has demonstrated that Zn dissolved well in molten ZnCl$_2$, which is attributed to a redox reaction between Zn and Zn$^{2+}$.[45, 46] As mentioned above, the Zn$^{2+}$ cation is a strong electron acceptor that can react with Zn to form lower valent Zinc cations (Zn$^+$ or Zn$_2^{2+}$). Therefore, the weakly bonded Zn atoms in Ti$_3$ZnC$_2$ were easily extracted from the two-dimensional A-site plane and dissolved in the ZnCl$_2$ molten salt. Meanwhile, according to reaction (7), the Cl$^-$ anions spontaneously intercalated into the A-site plane and bonded with the specific site in between Ti$_3$C$_2$ sublayers to form a more stable phase, i.e. Ti$_3$C$_2$Cl$_2$. The strong electronegativity of Cl results in the increasing valence of the Ti$_3$C$_2$ unit, which means that extra electrons were released. These were captured by Zn$^+$ or Zn$_2^{2+}$ ions and formed pure metal Zn (reaction (8)). Note that the valence state of Ti is also confirmed by the above XPS analysis (Fig. 3d). In other words, the formation mechanism of Ti$_3$C$_2$Cl$_2$ is comparable to the formation chemistry of HF etching on Ti$_3$AlC$_2$,[3] in which Zn$^{2+}$ and Cl$^-$ plays a similar role as H$^+$ and as F$^-$, respectively.

Table 2 shows the cleavage energies of the M-Zn atomic layer in four Zn-MAX phases. The M-Zn cleavage energies of Ti$_2$ZnN and V$_2$ZnC are theoretically predicted to be higher than that of Ti$_3$ZnC$_2$ and Ti$_2$ZnC, indicating a stronger M-Zn bond. The stronger bond indicates that a more severe etching condition than in the present work is required to remove the Zn atoms from Ti$_2$ZnN and V$_2$ZnC. The experimental results described above illustrated that at the same reaction condition, no corresponding MXCl$_2$ MXenes can be obtained. This fact is also in agreement with the HF etching chemistry that MAX phases with higher M-A bonding energy, such as V$_2$AlC and Nb$_2$AlC, requires longer times and higher HF concentrations to etch out the Al atoms than that of Ti$_2$AlC.[5] This phase formation, together with the following calculation results, explain the reason why Ti$_3$ZnC$_2$ and Ti$_2$ZnC could convert into Ti$_3$C$_2$Cl$_2$ and Ti$_2$CCl$_2$, whereas V$_2$ZnC and Ti$_2$ZnN were not converted into the corresponding Cl-MXenes.

Table 2 Calculated cleavage energy $E$ (J/m$^2$) of different M and Zn atomic layer.

|  | Ti$_3$ZnC$_2$ (Ti-Zn) | Ti$_2$ZnC (Ti-Zn) | V$_2$ZnC (V-Zn) | Ti$_2$ZnN (Ti-Zn) |
|---|---|---|---|---|
| $E$ | 1.552 | 1.576 | 1.774 | 1.754 |
| derivative MXene | Ti$_3$C$_2$Cl$_2$ | Ti$_2$CCl$_2$ | NA | NA |

NA: not available in present work.

**Summary and outlook**

In summary, a series of novel MAX phases ($Ti_3ZnC_2$, $Ti_2ZnC$, $Ti_2ZnN$, and $V_2ZnC$) as well as Cl-terminated MXenes ($Ti_3C_2Cl_2$ and $Ti_2CCl_2$) were synthesized by a replacement reaction, where the A element in traditional MAX phase precursors are replaced by Zn from $Zn^{2+}$ cation in molten $ZnCl_2$.

The formation of the Zn-MAX phases were achieved by a replacement reaction between $Zn^{2+}$ and Al, and subsequently the occupancy of Zn atoms to the A sites of the MAX phase. The formation of Zn-MAX phases indicates that such exchange mechanism between traditional Al-MAX phases and the late transition metal halides might be a general approach to synthesize some other unexplored MAX phases with functional A-site elements (such as the magnetic element Fe). Late transition metal halides, e.g., $ZnCl_2$, which have relatively low melting point and exhibit strong Lewis acidity in their molten state, seem to be ideal candidates for the replacement reaction. The acidic environment provided by the molten salts facilitates extraction of the Al atoms from the A atom plane in MAX phase at a moderate temperature. The generation of the volatile Al halides in turn, provides the driving force for the out-diffusion of the Al atom. Meanwhile, the liquid environment also facilitates the in-diffusion of replacement atoms, which finally promotes a thorough replacement reaction.

The formation of $Ti_3C_2Cl_2$ and $Ti_2CCl_2$ MXenes was achieved by a further exfoliation of $Ti_3ZnC_2$ and $Ti_2ZnC$ in $ZnCl_2$ molten salt. $Ti_2NCl_2$ and $V_2CCl_2$ was not obtained since their Zn-MAX phases has higher M-A bonding strength than $Ti_3ZnC_2$ and $Ti_2ZnC$. Significantly, this is also the first time that exclusively Cl-terminated MXenes were obtained through a non-Fluorine chemistry. The Cl-terminated MXenes are expected to be more stable than F-terminated MXenes which implicates promising applications such as energy storage. [7, 8] Actually, a few recent reports have indicated enhanced electrochemical behavior in Cl functionalized MXenes. [47, 48] The moderate molten-salt environment also warranty an viable and green chemistry for fabrication of MXenes that paths the way for their scale-up and even commercial application.

**Acknowledgement**s

This study was supported financially by the National Natural Science Foundation of China (Grant No. 21671195 and 91426304), and China Postdoctoral Science

Foundation (Grant No. 2018M642498). The Knut and Alice Wallenberg Foundation is acknowledged for support of the electron microscopy laboratory in Linköping, Fellowship grants and a project grant (KAW 2015.0043). P. O. Å. P, J. R. and J. L. acknowledge the Swedish Foundation for Strategic Research (SSF) for project funding (EM16-0004) and the Research Infrastructure Fellow RIF 14-0074. We also acknowledge support from the Swedish Government Strategic Research Area in Materials Science on Functional Materials at Linköping University (Faculty Grant SFO-Mat-LiU No. 2009 00971).

**Supporting Information.**

Weight change during the synthesizing process; Rietveld refinement of the XRD pattern; additional SEM images and EDS data; additional TEM figures; detailed XPS results; DFT calculation results; additional XRD results; phase diagrams of the Ti-Al-C and Ti-Zn-C systems.